\documentstyle[twoside,fleqn,espcrc2,epsf]{article}

\newcommand{\dslash}{\partial\hspace{-.075in}/}

\newcommand{\lsim}{\mathrel{\lower4pt\hbox{$\sim$}}
\hskip-12.5pt\raise1.6pt\hbox{$<$}\;}
\newcommand{\gsim}{\mathrel{\lower4pt\hbox{$\sim$}}
\hskip-12.5pt\raise1.6pt\hbox{$>$}\;}
\title{The neutrino ground state in a neutron star}

\author{Ken Kiers\address{High Energy Theory, Department of Physics\\ 
		Brookhaven National Laboratory,
		Upton, NY 11973-5000, USA}%
        \thanks{Talk presented at ``NEUTRINO 98'', Takayama, Japan, 
		June 4-9, 1998.  This contribution to the proceedings
		is excerpted from Ref.~\cite{kt}.}%
	\thanks{Address after September, 1998: Physics
		Department, Taylor University, 236 West
		Reade Ave., Upland, IN 46989, USA}
        and 
       Michel H.G. Tytgat\address{Service de Physique Th\'eorique,
		Universit\'e Libre de Bruxelles, CP225\\
		Bd du Triomphe, 1050 Bruxelles, Belgium}}
       
\begin{document}

\begin{abstract}

We address a recent claim that the stability of neutron
stars implies a lower bound on the mass of the neutrino.
We argue that the result obtained
by some previous authors is due to an improper summation
of an infrared-sensitive series and that a non-perturbative
``resummation'' of the series yields a finite and well-behaved
result.  The stability of neutron stars
thus gives no lower bound on the mass of the neutrino.
\end{abstract}

\maketitle

\section{INTRODUCTION}
\label{sec:intro}
In this talk we present a calculation of the interaction 
energy due to multi-body neutrino exchange in a neutron
star~\cite{kt}.

\begin{figure}[b]
\epsfbox[200 630 400 630]{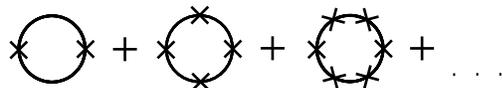}
\caption{\label{fig:pertexp} Perturbative expansion of the 
	shift in the neutrino ground state energy due to the
	presence of a neutron star.  Solid lines
	represent neutrino propagators and crosses represent
	insertions of the neutron density.}
\end{figure}

Consider the series shown in Fig.~\ref{fig:pertexp}, which represents the 
``self-energy'' of a neutron star due to neutrino exchange.  
In this figure the crosses represent insertions of the neutron
density.  It is not difficult to see that the term in 
this series with $k$ insertions of the neutron density 
scales approximately as~\cite{fb}
\equation
	W^{(k)} \sim \frac{C_k}{R}\left(\frac{G_F N}{R^2}\right)^k, 
\endequation
\noindent
where $N\sim10^{57}$ is the total number of neutrons in the 
star, $R\sim 10$ km is the radius of the star and $C_k$ is
a dimensionless numerical coefficient.  The thing which
is perhaps surprising in this expansion is that the
``expansion parameter'' $G_F N/R^2$ is of order $10^{12}$.
A direct summation of the terms in this series (but truncated
after $N$ terms) yields an enormous value for the interaction
energy~\cite{fb}.  This result has led some previous authors
to claim that neutrinos must have a mass of at least $0.4$ eV
in order to allow neutron stars to exist as stable 
objects~\cite{fb,fb2,fischxxx}.

Our approach to this problem is quite different.
We contend that the series represented in Fig.~\ref{fig:pertexp} is actually
infrared-divergent and must be ``resummed'' non-perturbatively
in order to yield a sensible result.  This approach was previously
advocated in Ref.~\cite{abada}.  We have performed the required
resummation and find that the apparent infrared divergence
is an artifact of the expansion and that the interaction energy is
finite and well-behaved~\cite{fn1}.  There is thus no lower bound on
the mass of the neutrino.  Along the way we also encounter some
interesting physics.  We demonstrate, for example, that the
ground state of the system actually contains a non-zero neutrino
number -- a result which was previously anticipated in Ref.~\cite{loeb}.
In our simple model for the density of the neutron star it is
straightforward to calculate both the energy and the neutrino
number of the ground state.

Before describing the calculation in more detail, let us
note that there have been several groups which have examined
various aspects of this 
problem~\cite{abada,smirnov1,recone,rectwo,penexxx,arafune}.
In particular, Arafune and Mimura~\cite{arafune}
have confirmed our asymptotic result using an analytical approximation.

\section{NEUTRINO GROUND STATE}
\label{sec:rev}

\subsection{Preliminaries}
Our goal is to calculate the shift in the neutrino ground state energy due
to the presence of the star.  This energy shift may be defined in
terms of the neutrino Hamiltonian $H_{(0)}$ in the presence (absence) of
the star as follows~\cite{julian}:
\begin{equation}
\label{shift}
W = \langle \hat 0 \vert  H \vert \hat 0 \rangle - \langle 
 0 \vert  H_0 \vert  0 \rangle .
\end{equation}
Here $\vert \hat 0 \rangle$ denotes the neutrino
ground state in the presence of the star, while $\vert 0 \rangle$
denotes the usual matter-free vacuum state.  As we have already
alluded, the state $\vert \hat 0 \rangle$ contains in general a 
non-zero neutrino number (i.e., it is ``charged'').
Note that the expression in Eq.~(\ref{shift}) is a formal, 
ultraviolet-divergent quantity which needs to be renormalized.  
This renormalization may be done quite easily using the usual techniques.

In order to proceed, it is convenient to introduce an
effective Lagrangian for the neutrino field.  After integrating
out all of the other particles in the theory, one 
obtains~\cite{mann,abada,ken2}
\begin{equation}
\label{efflag}
{\cal L}_{\rm eff} = 
  \overline{\psi}_{L} \left[i\dslash 
        + \alpha \gamma^0\right]\,\psi_{L}
\end{equation}
where $\psi_L = \frac{1}{2}(1-\gamma_5) \psi$, and where
\begin{equation}
\alpha(\vec{x}) = G_F\rho_n(\vec{x})/ \sqrt{2}\, \sim 20\,\,{\rm eV}
\end{equation}
is the  electroweak  potential 
induced by the finite neutron density 
($\rho_n \approx 0.4\,\, \mbox{\rm fm}^{-3}$ 
in a typical neutron star).  This potential is
identical to the one which is usually considered 
in the well-known Mikheyev-Smirnov-Wolfenstein (MSW) effect~\cite{msw}.
Note that the potential term in (\ref{efflag}) resembles
a position-dependent chemical potential, so that it is not
at all surprising that the ground state of the system has
a non-zero neutrino number.

It is straightforward to derive the following
perturbative expansion for $W$ in terms of the potential $\alpha(\vec{x})$
and the neutrino propagator $G_0(\vec{x},\vec{x}^\prime;\omega)$~\cite{kt}:
\begin{eqnarray}
\label{exp2}
	W &  = & {1\over 2 \pi i} \sum_{k = 1}^\infty \, {1 \over k}\,
 \int_C d\omega\, 
 \mbox{\rm Tr}_{\bf x} \left  [{\alpha  G_0(\omega)}\right ]^k \\
	& \equiv & \sum_k W^{(k)} .
	\label{eq:pertexp}
\end{eqnarray}
All of the odd terms in this expansion
disappear so that this series corresponds precisely 
to that which is represented diagrammatically in Fig.~\ref{fig:pertexp}.
One aspect of the perturbative expansion which is useful
is that it neatly isolates
the ultraviolet divergence in $W$.  In fact, the only ultraviolet
divergent term in (\ref{eq:pertexp}) is that with $k=2$.  This
term is related to the vacuum polarization of the $Z$ in the
complete theory.  A final note concerning this expansion is 
that while the terms with $k\geq 4$ are separately finite,
their sum is infrared-sensitive and is in fact ill-defined
for ``large'' stars [$\alpha R \gg {\cal O}(1)$].  The non-perturbative
``resummation'' of these terms is the main goal of our calculation.

\subsection{Comparison point}

It is useful to compare the results which we will describe with
results which were obtained previously in the literature.
The perturbative expansion of $W$ has been considered in 
Refs.~\cite{fb,fb2,fischxxx}, where it was found that
already by the eighth term in the expansion the interaction 
energy exceeded the gravitational binding energy of the star.
After summing up the entire series [which in the approach
of Refs.~\cite{fb,fb2,fischxxx} was actually a truncated sum,
not an infinite series as we have in Eq.~(\ref{eq:pertexp})],
it was found that the interaction energy exceeded the rest
mass of the universe.  It was argued that the only way to regulate
the sum was to give all neutrino flavours a mass of at least $0.4$ eV.

Our philosophy in this matter is that the perturbative
expansion is simply outside of its radius of convergence
when $\alpha R \gg {\cal O}(1)$ and that the series needs
to be resummed using a non-perturbative approach~\cite{abada}.
We cannot consider a realistic neutron star (i.e., with 
$\alpha R\sim G_F N/R^2\sim 10^{12}$) using our
numerical approach, but it is actually sufficient to
restrict our analysis to $\alpha R\leq {\cal O}(100)$, since 
for $\alpha R \sim {\cal O}(1)$ we already observe
a ``cross-over'' to the non-perturbative regime.
A clear signal of this cross-over is that the ground
state obtains a non-zero charge.  (The charge of the ground state is 
exactly zero to any finite order in perturbation theory.)

Consider the comparison point $\alpha R = 20$.
This point could correspond to a tiny ``star'' with 
a realistic neutron density ($\alpha = 20$ eV), but with a
tiny radius ($R = 2\times 10^{-5}$ cm).  In this case
the truncated sum in Ref.~\cite{fb2} 
gives $\sum_{k=4}^N W^{(k)}\sim 10^{66}$~eV.  By way of comparison,
our non-perturbative resummation gives $\sum_{k=4}^N W^{(k)}\sim -2.3$ keV.

\subsection{Phase shift formulas}

An exact, non-perturbative expression for the interaction energy $W$
is given by the following expression due to Schwinger~\cite{julian}
\equation
	W = \frac{1}{2\pi} \sum_{l=0}^{\infty} (2l+2) \int_{0}^\infty
		d\omega \; \left[\delta_l(\omega)+\delta_l(-\omega)
		\right] . \label{wformal}
\endequation
\noindent
Here $\delta_l(\omega)$ is the scattering phase shift for a neutrino
incident on the star and $l$ labels the orbital angular momentum.
A similar expression may be obtained for the ``charge'' of the ground
state:
\begin{eqnarray}
	q\!\!\!\!&=\!\!\!\!&
		-\frac{1}{2\pi}\sum_{l=0}^{\infty} (2l+2) \nonumber \\
		& & \times \int_{0}^\infty d\omega \;
		\left[\frac{\partial}{\partial\omega}\delta_l(\omega)-
		\frac{\partial}{\partial\omega}\delta_l(-\omega)\right] .
	\label{qformal}
\end{eqnarray}
The factor $(2l+2)\equiv (2 j +1)$ is the degeneracy factor for a given
energy $\omega$ and total angular momentum $j$.

The beauty of the above expressions for $W$ and $q$ are that they
are valid for any value of $\alpha R$.  Note, however, that 
Eqs.~(\ref{wformal}) and (\ref{qformal})
are still formal ultraviolet-divergent expressions 
which need to be renormalized.

\section{NUMERICAL EVALUATION}

Let us first choose a density profile for the neutron star.
Since the quantities which are of interest to us (i.e., $W$ and $q$)
are only sensitive to the gross features of the star, it is
convenient to choose a very simple -- albeit unrealistic -- density
profile:
\begin{equation}
	\alpha(r) = \alpha \theta(R-r) .
\end{equation}
Recall that $\alpha = G_F \rho_n(\vec{x})/\sqrt{2} \sim 20$~eV in
a neutron star.  In this very simple model it is straightforward to
obtain closed expressions for the scattering phase shifts, which
simplifies our numerical work considerably.

We may now renormalize $W$ and $q$.  Since our 
model is renormalizable the ultraviolet divergences in $W$ and $q$
are confined to the first few terms in the perturbative expansion.
These terms may be isolated by Taylor-expanding the phase shift
formulas in $\alpha R$.  [Note that by Taylor-expanding 
in $\alpha R$ we recover the perturbative expansion defined in
Eq.~(\ref{exp2}).]  The procedure is then as follows: (i) Taylor
expand $W$ and $q$ in order to isolate the divergent terms; (ii)
subtract out the divergent terms; (iii) regularize and renormalize
the divergent terms using conventional methods; (iv) add the finite, 
renormalized terms back in.  This procedure yields the following
expression for the renormalized energy:
\begin{equation}
	W_{\rm ren} = W_{\rm ren}^{(2)} + W^{(4+)} ,
	\label{wrendef}
\end{equation}
where
\begin{equation}
	W^{(4+)}\!\!= \!\frac{1}{2\pi} \sum_{l=0}^{\infty} 
			(2l+2)\!\!\int_{0}^\infty
		\!\!\!\!d\omega \!\left[\overline{\delta}_l(\omega)\!+\!
			\overline{\delta}_l(-\omega)\right]\! ,
	\label{eq:w4plus}
\end{equation}
and where $\overline{\delta}_l(\omega) \equiv \delta_l(\omega)
- \delta_l^{(2)}(\omega)$,
with $\delta_l^{(2)}(\omega)$ being the second term in the Taylor
expansion of $\delta_l(\omega)$.  $W_{\rm ren}^{(2)}$ is essentially
the vacuum polarization of the $Z$ in the full theory (but convoluted
over the neutron star) and corresponds to the first term in the
diagrammatic expansion in Fig.~\ref{fig:pertexp}.  
This term has been discussed in detail in Ref.~\cite{kt} and will not
be considered further here.
The second term in Eq.~(\ref{wrendef}), $W^{(4+)}$,
is of particular interest to us since it
represents the non-perturbative ``resummation'' of the
terms with four or more insertions of the neutron density
in the diagrammatic expansion shown in Fig.~\ref{fig:pertexp}.
As we have noted, a direct summation of these terms
does not lead to a sensible result when $\alpha R >{\cal O}(1)$.
The resummed result in Eq.~(\ref{eq:w4plus}), however, is always
well-defined and leads to a well-behaved and sensible result.

We may similarly obtain an expression for the renormalized charge:
\begin{equation}
	q_{\rm ren} = \frac{1}{2\pi} \sum_{l=0}^{\infty} (2l+2)
			\left[\delta_l(0^+)-\delta_l(0^-)\right] .
	\label{eq:qren}
\end{equation}
In order to calculate the charge, then, we need only know 
the scattering phase shifts at the origin.

\subsection{Small $\alpha R$}

For small $\alpha R$ the ``resummed'' energy, $W^{(4+)}$, 
is well-approximated by the leading term in the Born expansion.  
We have checked this explicitly by calculating $W^{(4+)}$ directly using the
phase shift formula and by Taylor-expanding the phase shifts to fourth
order and (analytically) integrating the resulting expressions.
The result which we obtain is
\begin{equation}
	W^{(4+)} \simeq W^{(4)} \simeq -0.00115 \frac{(\alpha R)^4}{R} .
\end{equation}
For small $\alpha R$ the phase shifts are zero at the origin, so
that the ground state of the system is uncharged [see Eq.~(\ref{eq:qren})].

\subsection{Larger $\alpha R$}

As $\alpha R$ is increased, there is a critical
point beyond which it becomes energetically favourable 
for the neutrinos to ``condense''; thus, for 
$\alpha R$$>$$\left.\alpha R\right|_{\rm crit}$, the ground state of the
system carries a non-zero neutrino number.
There are in fact an infinite number
of points at which the charge of the
ground state changes discontinuously as $\alpha R$ is increased.
In our simple model for the density profile of the neutron
star there is a correspondingly simple condition for a new
charge to be added to the ground state:
\begin{equation}
	j_l(\alpha R) = 0 .
\end{equation}
In the quantum mechanical scattering
problem, these values of $\alpha R$ correspond to
the points at which a resonance crosses from the positive to
the negative energy continuum~\cite{kt}.

\begin{figure}[htb]
\hbox to\hsize{\hss\epsfxsize=8cm\epsfbox[134 100 449 287]{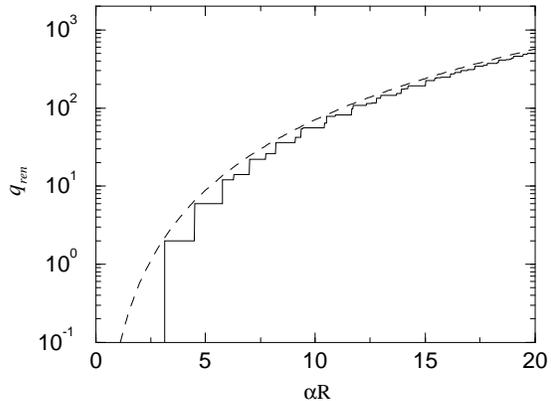}\hss}
\caption{Plot of the charge as a function of $\alpha R$.
The solid curve gives the exact charge
and the dashed curve gives the charge expected for a system 
with chemical potential $\mu = \alpha$.}
\label{fig:qren}
\end{figure}

Figure~\ref{fig:qren} shows a plot of the renormalized charge
as a function of $\alpha R$.  The solid curve gives the exact charge, 
which has periodic jumps, and the dashed curve gives the charge 
expected in the large volume limit
for a system with chemical potential $\mu = \alpha$:
$q_{\rm cond} = 2(\alpha R)^3/(9\pi)$.
Clearly, as $\alpha R$ gets large the exact result tends to this limit.
The critical point -- at which the ground state becomes charged
and beyond which we should not trust the perturbative expansion -- is
seen to occur at $\alpha R = \pi$ in this model.  This is the 
first zero of $j_0(\alpha R)$.

\begin{figure}[hbt]
\hbox to\hsize{\hss\epsfxsize=8cm\epsfbox[140 100 449 488]{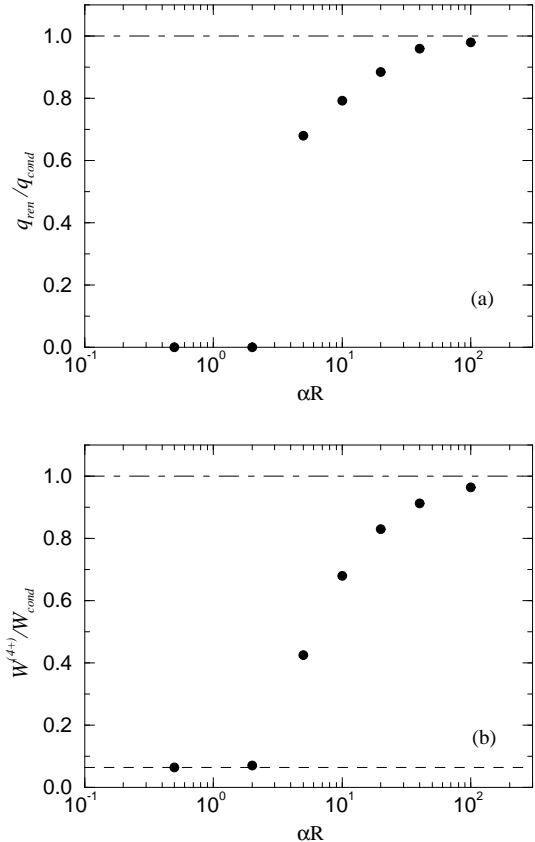}\hss}
\vspace{.2in}
\caption{Normalized plots of (a) the charge and (b) the 
energy as a function of $\alpha R$.  The dots give the results of our
exact (non-perturbative) calculations.}
\label{fig:chargeen}
\end{figure}

Figure~\ref{fig:chargeen} shows plots of both the charge
and the energy as functions of $\alpha R$.  Both the charge and the
energy are normalized to the values expected for a 
``condensate'' in the large-volume limit.  For small $\alpha R$
the charge is zero and the energy is well-described by 
the leading term in the Born expansion [which is shown by the lower
dashed line in Fig.~\ref{fig:chargeen}(b)].  At $\alpha R = \pi$
the ground state becomes charged and perturbation theory
breaks down.  As $\alpha R$ is increased, both the charge and the
energy tend toward the values expected for a large system
with chemical potential $\mu = \alpha$.  [Note that 
$W_{\rm cond} = -\alpha^4R^3/(18\pi)$.]
The asymptotic trend apparent in Fig.~\ref{fig:chargeen}
has recently been confirmed by Arafune and Mimura using an
analytical approximation~\cite{arafune}.

The comparison point which was singled out above ($\alpha R = 20$)
is included in the points shown in Fig.~\ref{fig:chargeen}.
A previous (perturbative) calculation of the energy associated with this 
point yielded an enormous
value~\cite{fb2}, while our non-perturbative
approach yields a very small and innocuous value.

\section{CONCLUSIONS}

It has been argued in 
Refs.~\cite{fb,fb2,fischxxx} that the interaction energy due
to the exchange of massless or very light neutrinos in a neutron
star would be enormous and would destabilize the star.  We have
addressed this claim by performing an explicit non-perturbative
calculation of the interaction energy.  We find that, once 
properly resummed, the interaction energy is a finite and well-behaved
quantity which is far too small to have any effect on the fate
of a neutron star.  Thus, contrary to previous claims, the stability
of neutron stars places no lower bound on the mass of the neutrino.

\vspace{.1in}
We are indebted to R. Jaffe for many insightful and helpful 
conversations during the course of this work.
This research was supported in part by the U.S. Department of Energy
under contract number DE-AC02-76CH00016.
K.K. is also supported in part by the Natural Sciences and
Engineering Research Council of Canada.  K.K. also wishes to thank
the conference organizers for partial support.


\begin{thebibliography}{99}
\bibitem{fn1}
In order to simplify our calculation,
we consider only the case $m_\nu = 0$, but by extension our
result is also valid for very light -- but not strictly massless -- neutrinos.
\end{thebibliography}
\end{document}